\documentclass[%
 reprint,
 amsmath,amssymb,
 aps,
pre,
]{revtex4-1}

\usepackage{graphicx}% Include figure files
\usepackage{dcolumn}% Align table columns on decimal point
\usepackage{bm}% bold math
\begin{document}

\preprint{APS/123-QED}

\title{Percolation thresholds of two-dimensional continuum systems of rectangles}% Force line breaks with \\

\author{Jiantong Li}
\email{jiantong@kth.se}
\author{Mikael \"Ostling}%
\affiliation{%
 KTH Royal Institute of Technology, School of Information and Communication Technology, Electrum 229, SE-164 40 Kista, Sweden
}%


\date{June 4, 2013}% It is always \today, today,

\begin{abstract}
The present work introduces an efficient Monte Carlo algorithm for continuum percolation composed of randomly-oriented rectangles. By conducting extensive simulations, we report high-precision percolation thresholds for a variety of homogeneous systems with different rectangle aspect ratios. This work verifies and extends the excluded area theory. It is confirmed that percolation thresholds are dominated by the average excluded areas for both homogeneous and heterogeneous rectangle systems (except for some special heterogeneous systems where the rectangle lengths differ too much from one another). In terms of the excluded areas, generalized formulae are proposed to effectively predict precise percolation thresholds for all these rectangle systems. This work is therefore helpful for both practical applications and theoretical studies concerning relevant systems.
\begin{description}

\item[PACS numbers]
64.60.ah

\end{description}
\end{abstract}

\pacs{Valid PACS appear here}% PACS, the Physics and Astronomy
\maketitle

%\tableofcontents

\section{\label{sec:I}INTRODUCTION}

Over the past decade, layered two-dimensional (2D) nanomaterials, such as graphene, MoS$_2$, WS$_2$, and boron nitride (BN) \cite{Ref1, Ref2, Ref3,Ref4, Ref5}, have attracted great interests in the fields of material science, electronics, medicine, biology, and so forth. Many applications require these 2D materials to be integrated into percolating systems to acquire certain functions \cite{Ref2, Ref5, Ref6, Ref7}. In general, a simple and representative model for these systems should be a percolation system comprising randomly-oriented rectangles \cite{Ref8} (an example is shown in Fig. \ref{fig:1}). As a matter of fact, the interest in rectangle percolation is not limited to practical applications for these emerging 2D nanomaterial systems. It can date back to 1980s for theoretical studies concerning the excluded area (volume) \cite{Ref9,Ref10}.

Unfortunately, rectangle systems suffer from some fundamental problems not yet addressed. For example, the percolation thresholds are not known even for the simplest 2D homogeneous rectangle systems, despite the already known thresholds in high precision for similar 2D continuum systems composed of random disks \cite{Ref11,Ref12}, squares \cite{Ref12,Ref13} and sticks \cite{Ref12,Ref14}. The unknown percolation threshold for rectangle systems has actually also caused difficulty in verifying in these systems the excluded area theory \cite{Ref9,Ref15} which assumes that the percolation threshold $N_c$ is inversely proportional to the average excluded area $A_r$:
\begin{equation}
 N_c\propto A_r^{-1}
 \label{eq:one},
\end{equation}
where $N_c$ is the critical number density of rectangles at percolation. The excluded area (volume) is defined as the minimum area (volume) around an object into which the center of another similar object cannot enter in order to avoid the overlapping of the two objects \cite{Ref9}. In homogeneous systems, $A_r$ for randomly-oriented rectangles of length $l$ and width $w$ (and then the aspect ratio $r=l/w$) is \cite{Ref9}
\begin{equation}
\begin{split}
 A_r &=2lw(1+4\pi^{-2})+2(l^2+w^2)\pi^{-1}\\
       &=2l^2[(1+4\pi^{-2})r^{-1}+\pi^{-1}(1+r^{-2})]
 \end{split}
 \label{eq:two}.
\end{equation}

In addition, practical applications may show more interest in heterogeneous rectangle systems which comprise various types of rectangles of different $r$ and/or different $l$. A typical example is the composites of 2D layered materials (low-$r$ rectangles) and carbon nanotubes (high-$r$ rectangles) \cite{Ref3, Ref16, Ref17}. These systems have even not been considered in the traditional excluded area theory \cite{Ref9}.

In order to meet the requirements for practical applications, and verify and extend the excluded area theory, Monte Carlo simulations are performed in this work to explore percolation thresholds for a variety of 2D rectangle systems. 

\section{\label{sec:II}SIMULATION ALGORITHM}
\begin{figure}
\includegraphics[scale=0.5]{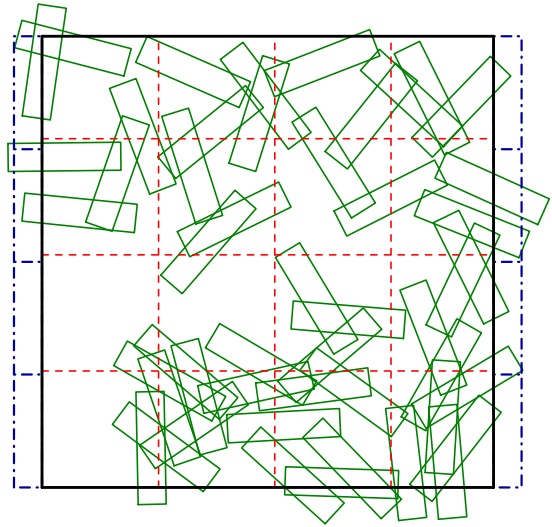}% Here is how to import EPS art
\caption{\label{fig:1}(Color online) An example of a homogeneous FBC rectangle percolation (produced by our simulation program). Here the rectangles (olive) are of length $l = 1$, width $w = 0.25$, and aspect ratio $r = 4$. The system (of size $L = 4$) is defined by the four bold (black) boundaries. For convenience in employing the Newman-Ziff algorithm, each of the two opposite (left and right) boundaries is represented by $L$ (= 4 here) vertically-oriented intersecting auxiliary rectangles [dash dotted (blue) rectangles]. The system spans (percolates) if two opposite auxiliary rectangles are connected by the edges of normal rectangles, like the case in this figure. The whole systems are virtually divided by a number of subcells, as shown by the dashed (red) lines. All subcells are of the same side length $\sqrt{l^2+w^2}$, except those in the topmost row or rightmost column. Each rectangle belongs to a subcell in which its center lies. Each auxiliary rectangle belongs to its nearest subcell. Note in general, subcell side length is larger than the rectangle length. }
\end{figure}

Previously, by combining the fast Newman-Ziff algorithm \cite{Ref18} with the subcell concept \cite{Ref19}, we have developed an efficient algorithm for stick percolation \cite{Ref14}. In brief, during the simulations, each stick is registered into a subcell where its center lies so that it is only necessary to check the connectivity between two sticks belonging to the same and 8 neighboring subcells. And each cluster which comprises connecting sticks is stored by a tree structure \cite{Ref18} so that the status of a cluster can be readily updated when a new stick is added into the system. This algorithm provides comparable efficiency to those for lattice percolation \cite{Ref14,Ref18}. In fact, it can be readily generalized to any continuum percolation systems, including certainly the rectangle systems here. Like a stick \cite{Ref14}, a rectangle can also be stored simply through a point (its center) and an angle (its orientation). Therefore, the transfer from stick systems to homogeneous rectangle systems only needs two main modifications. One is that the subcell length should be set as the diagonal length of the rectangles, that is, $\sqrt{l^2+w^2}$ or $l\sqrt{1+r^{-2}}$, as shown in Fig. \ref{fig:1}. The other is a different technique should be employed to check the connectivity between two rectangles in the same and neighboring subcells \cite{Ref14}. In this work, we adopt the Cohen-Sutherland (CS) algorithm, a famous line clipping algorithm in computer graphics \cite{Ref20}. It can efficiently determine whether a line segment is visible (or partially visible) in a viewport (a horizontally-oriented rectangular window). In detail, the CS algorithm divides the 2D space into 9 regions in terms of the viewport. Each region has a 4-bit code, and one can readily determine the regions of a line segment’s end points by bitwise operations. A line segment is “trivially accepted” (completely visible) if both of its end points are inside the viewport region, or “trivially rejected” (completely invisible) if the two end points are on the same side of the viewport. Otherwise, one end point which is outside the viewport is replaced with the intersection point between its nearby (extended) viewport edge and the line segment. The process is repeated until the line segment is “trivially accepted” or “trivially rejected”. Note the CS algorithm \cite{Ref20} attempts to determine the visible portion of a line segment, while rectangle percolations are only interested in its visibility. Therefore, we can further simplify the CS algorithm by relieving the “trivially accepted” conditions: A line segment is “trivially accepted” if at least one of its end points is inside the viewport or one end point is over and the other is under the viewport. 

In our simulations, to check the connectivity between two rectangles, we set one as the reference rectangle and the other as the test rectangle, and translate and rotate the space so that the reference rectangle becomes the viewport (a horizontally-oriented rectangle centered at the origin). Since the target for the reference rectangle is always the viewport, the translation and rotation are only necessary for the test rectangle. Then, for each side of the test rectangle, which is a line segment, we use the CS algorithm to check whether it is visible in the viewport (reference rectangle). Once one side is found visible, one can conclude that the two rectangles connect (intersect); otherwise, if none of the four sides is visible, the two rectangles do not connect. In order to further improve efficiency, prior to the CS process, we carry out a screening: If the center distance of two rectangles is larger than the sum of their half diagonal lengths, they are impossible to intersect and the CS process is not needed.

For heterogeneous systems which may comprise rectangles of different aspect ratios $r$ and/or different lengths $l$, the algorithm works in a similar way. In principle, only three steps need modification. First, the subcell length should be set as the largest diagonal length among all the rectangles. Second, during checking the pair connectivity, the reference rectangle should be the one of larger area $S = lw$. Otherwise, if a large test rectangle encloses a small reference rectangle (viewport), it might be “trivially rejected” by mistake by the CS algorithm. Third, for every newly-generated rectangle, an additional random number $x$ $(0 \le x \le 1)$ is allocated to determine its dimensions. A rectangle has length $l_i$ and aspect ratio $r_i$ if its $x$ satisfies $\sum\nolimits_{j=1}^{i-1}x_j<x\le\sum\nolimits_{j=1}^{i}x_j$, supposing the heterogeneous system comprises $n$ types of rectangles, and the $j$th type of rectangles have length $l_j$, aspect ratio $r_j$ and number fraction $x_j$ $(1 \le j \le n, \sum\nolimits_{j=1}^{n}x_j=1 )$.
Monte Carlo simulations (see Refs.\cite{Ref14,Ref18} for the detailed Newman-Ziff procedure) based on the algorithm above can produce the spanning probability for any system (either homogeneous or heterogeneous) as a function of the number (integers) of rectangles. Its convolution with the Poisson distribution \cite{Ref14} gives the spanning probability $R(N,L)$ for a system of size $L$ at {\it any arbitrary} rectangle number density $N$. In this work, all the systems are with free boundary conditions (FBCs) and the size $L$ is measured in units of $l_\textrm{max}$, the largest rectangle length. For simplicity, we set $l_\textrm{max}\equiv1$. From the known universal finite-size scaling $[R(N_c,\infty)=0.5 ]$ and $1/L$ corrections for the spanning probability of an FBC percolation system with square boundary \cite{Ref14,Ref21,Ref22}, one can expect 
\begin{equation}
 N_{0.5}(L)-N_c\sim L^{-1-1/v}  % \sim = ~
 \label{eq:three},
\end{equation}
where $N_{0.5}(L)$ is a percolation threshold estimate \cite{Ref14,Ref21} and defined through $[R(N_c,L)=0.5 ]$, and $v = 4/3$ is the critical correlation-length exponent. From Eq. (\ref{eq:three}), a high-precision threshold $N_c$, i.e., the thermodynamic limit ($L\rightarrow\infty$ ) value, can often be extracted for any percolation systems \cite{Ref14,Ref21,Ref23,Ref24}. 
\section{\label{sec:III}RESULTS AND DISCUSSION}
In this section, first, the performance of our algorithm is compared with other algorithms (Sec. \ref{sec:IIIA}). Then, simulation results and verification of the excluded area theory are presented and discussed for homogeneous rectangle systems (Sec. \ref{sec:IIIB}). Finally, the traditional excluded area theory is extended to heterogeneous rectangle systems (Sec. \ref{sec:IIIC}).
\subsection{\label{sec:IIIA}Performance of the algorithm}
In order to demonstrate the efficiency of the present algorithm (especially the integration of CS algorithm), we compare its performance with two other algorithms. Both of these two algorithms are still based on the combination of the subcell algorithm with Newman-Ziff procedure, but adopt different algorithms to check the connectivity between two rectangles: (i) The first one adopts “Edge-traverse” algorithm which tests whether any edge of the test rectangle intersects at least one edge of the reference rectangle \cite{Ref13}. The connectivity between two rectangle edges (actually line segments) is checked by the Pike and Seager’s bonding criterion of two sticks (also line segments) \cite{Ref25}, i.e., two sticks overlap if and only if for every stick, the distance between its center and the intersection point (of the corresponding lines for the two sticks) is no longer than its half-length. (ii) The second one employs another famous line clipping algorithm, the Liang-Barsky (LB) algorithm \cite{Ref20,Ref26}.
\begin{table}
\caption{\label{tab:table1}Average CPU time in hours consumed for each of the three algorithms discussed in the main text for identical homogeneous rectangle systems of $L = 128$. The average is over 20 independent batches of simulations and the uncertainty is the standard deviation. }
\begin{ruledtabular}
\begin{tabular}{lcc}
Algorithm &\mbox{$r =1$, $10^5$ runs}&\mbox{$r =10$, $10^4$ runs}\\
\hline
CS	& 0.796(14)	& 0.695(43) \\
Edge-traverse &	0.976(12)	& 0.948(42) \\
LB &	0.855(13) &	0.788(38)\\
\end{tabular}
\end{ruledtabular}
\end{table}

For clarity, hereafter the above three algorithms are denoted as “CS”, “Edge-traverse” and “LB”, respectively. Their performance has been tested via C++ codes on a server equipped with Intel(R) Xeon(R) CPU E5430 (2.66GHz). For each algorithm, we conducted 20 batches of Monte Carlo simulations for homogeneous rectangle systems of $L = 128$ for $r = 1$ and $r =10$, respectively. Each batch has one independent random seed and contains $10^5$ (for $r = 1$) or $10^4$ $(r = 10)$ simulation runs. The whole set of random seeds are identical for all three algorithms. As a result, the three algorithms produce identical simulation results (spanning probability functions), but consume significantly different CPU time. Table \ref{tab:table1} lists the average CPU time consumed for one batch. It is clear that the line clipping (CS and LB) algorithms are evidently faster than the Edge-traverse algorithm. In particular, the CS algorithm is ~20\% faster than the Edge-traverse algorithm for $r = 1$. With increasing $r$ to 10, the CS algorithm is even more efficient (~30\% faster than the Edge-traverse algorithm). Since the Edge-traverse algorithm has already been optimized in this work, the 20\%-30\% efficiency improvement of the CS algorithm should be thought as a considerable progress.

It is interesting to note that in the general case of line clipping, the LB algorithm has proven better efficiency than the CS algorithm \cite{Ref26}. For rectangle percolation, however, the alleviation of conditions for “trivial acceptance” increases the probability of a rectangle edge to be “trivially accepted” at the first test, so that the CS algorithm becomes more efficient \cite{Ref20}.

\subsection{\label{sec:IIIB}Homogeneous rectangle systems}	
\begin{figure}
\includegraphics[scale=0.3]{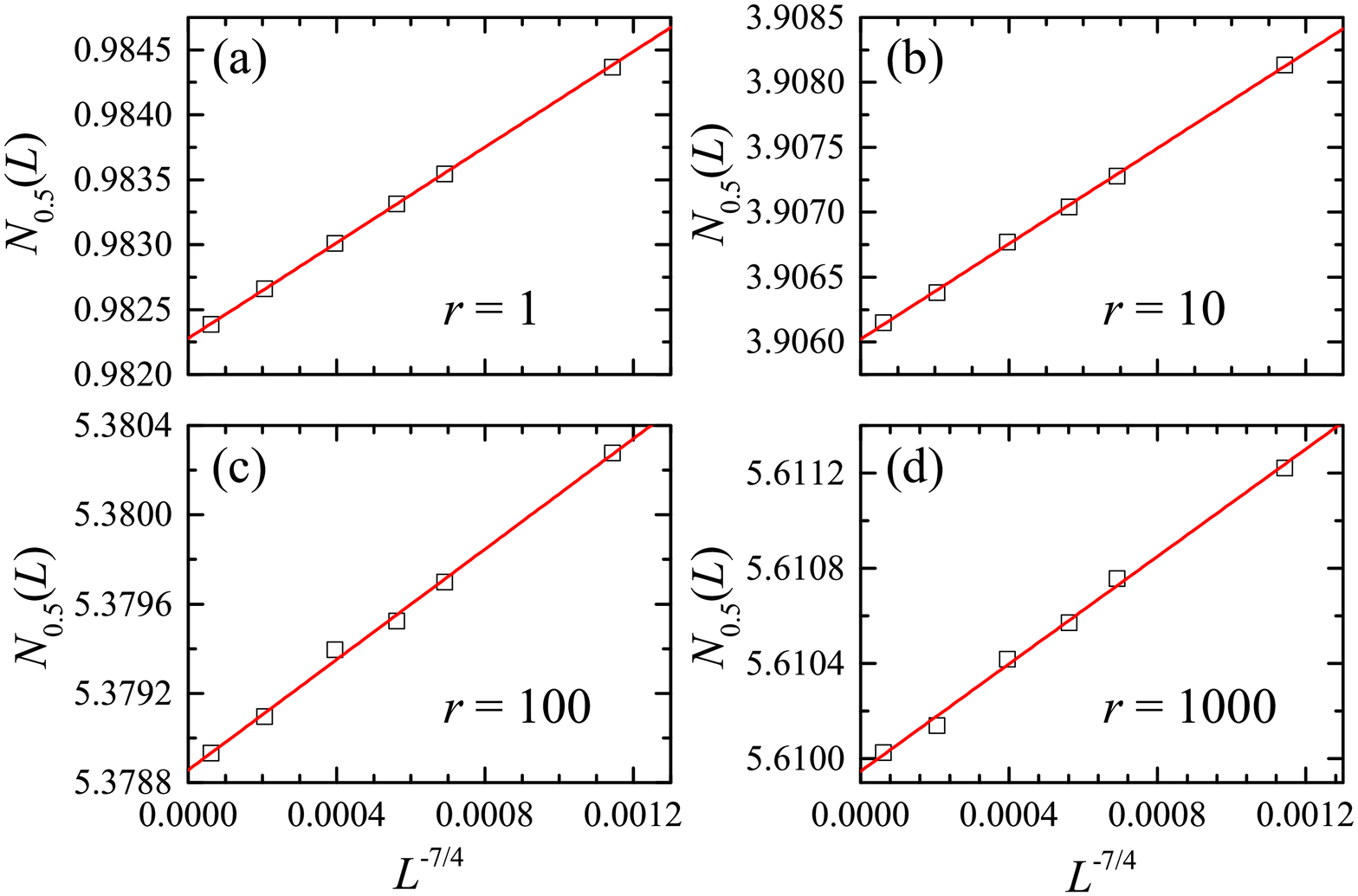}% Here is how to import EPS art
\caption{\label{fig:2}(Color online) Plots of $N_{0.5}(L)$ against $L^{-1-1/v}$ for homogeneous rectangle systems with (a) $r =1$, (b) $r =10$, (c) $r=100$, and (d) $r=1000$.}
\end{figure}

The CS algorithm is thereby employed to explore precise percolation thresholds for homogeneous rectangle systems. For each class of rectangle systems with the same $r$, we determine $N_{0.5}(L)$  for different sizes $L =$ 48, 64, 72, 88, 128 and 256, each of which is based on $>10^7$ Monte Carlo samples. It is found that for all $r$ studied in this work, $N_{0.5}(L)$ follows Eq. (\ref{eq:three}) quite well (adjusted $R^2 > 0.99$ for all the fittings; see Ref. \cite{Ref27} for definition of adjusted $R^2$). Some examples are shown in Fig. \ref{fig:2}. Then from linear fitting by Eq. (\ref{eq:three}) to these $N_{0.5}(L)$, we obtain high-precision values \cite{Ref23} for $N_c$ for a variety of $r$. As listed in Table \ref{tab:table2}, the uncertainty for all $r$ is less than 0.0001, which is given by the half-width of the 95\% confidence interval from linear fitting (regression) by Eq. (\ref{eq:three}). Note for $r = 1$, the rectangle systems actually turn to the square systems, and the value of $N_c$ is consistent with other work  \cite{Ref12,Ref13}. In contrast, for $r = \infty$, they degrade to stick systems, and the value of $N_c$ is taken from our previous work \cite{Ref14}. 

\begin{table}
\caption{\label{tab:table2}Percolation thresholds $N_c$ obtained from Monte Carlo simulations, their product with the exclude areas, $N_cA_r$, and the corrected product $(N_c-N_{c0})A_r$ for rectangle systems with different $r$. $N_{c0} = 0.20$. For $r = \infty$, $N_c$ is taken the value of stick systems \cite{Ref14}.}
\begin{ruledtabular}
\begin{tabular}{cccc}
\mbox{$r$} & \mbox{$N_c$} & \mbox{$N_cA_r$} & \mbox{$(N_c-N_{c0})A_r$} \\
\hline
1 &	0.982278(14)&	4.011&	3.195\\
1.5	& 1.425745(29)&	3.982&	3.424\\
2&	1.786294(26)&	3.932&	3.492\\
3&	2.333491(22)&	3.837&	3.508\\
4&	2.731318(30)&	3.767&	3.491\\
5&	3.036130(28)&	3.717&	3.472\\
6&	3.278680(19)&	3.681&	3.457\\
7&	3.477211(52)&	3.655&	3.445\\
8&	3.643137(24)&	3.635&	3.436\\
9&	3.784321(41)&	3.621&	3.429\\
10&	3.906022(37)&	3.609&	3.425\\
12&	4.105670(38)&	3.594&	3.418\\
15&	4.329848(28)&	3.580&	3.415\\
18&	4.495767(41)&	3.573&	3.414\\
20&	4.584535(54)&	3.570&	3.414\\
30&	4.878091(59)&	3.566&	3.420\\
50&	5.149008(20)&	3.569&	3.430\\
100&	5.378856(60)&	3.576&	3.443\\
1000&	5.609947(60)&	3.587&	3.459\\
$\infty$ &	5.637263(11)&	3.589&	3.461\\
\end{tabular}
\end{ruledtabular}
\end{table}

The product $N_cA_r$ is also listed Table \ref{tab:table2}. For all $r$ ranging from 1 to $\infty$, this product retains within a narrow region between 3.5 and 4.0. The region $3.5 < N_cA_r  \le 4.0$ is within, and much smaller than, the limits of $3.2 \le N_cA_r  \le 4.5$ predicted earlier \cite{Ref10}. It can thereby be roughly verified that, the excluded area theory \cite{Ref9} applies to rectangle systems. 

\begin{figure}
\includegraphics[scale=0.3]{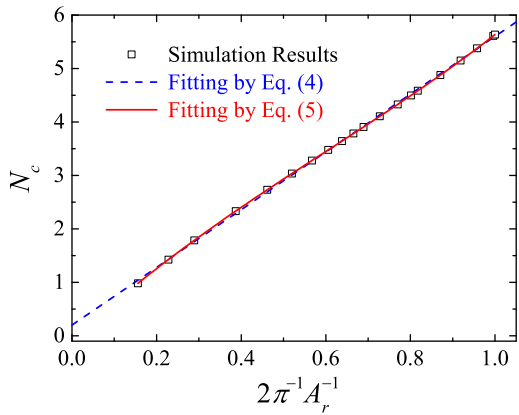}% Here is how to import EPS art
\caption{\label{fig:3}(Color online) Plots of $N_c$ against $2\pi^{-1}A_r^{-1}$ for homogeneous rectangle systems with $r$ ranging from 1 to $\infty$. Note all the systems are measured by their rectangle length, i.e., $l =1$. The simulation data (open squares) are also listed in Table \ref{tab:table2}. The dashed (blue) and solid (red) curves are fittings by Eqs. (\ref{eq:four}) and (\ref{eq:five}), respectively.}
\end{figure}

For more precise analysis, we plot all $N_c$ against the reciprocal of the normalized excluded area, $A_r^{-1}/A_\infty^{-1}$ or $2\pi^{-1}A_r^{-1}$  [note from Eq. \ref{eq:two}, $A_\infty=2\pi^{-1}\approx0.6366$  ]. As shown in Fig. \ref{fig:3}, the plot exhibits excellent linearity. However, the intercept is not at the origin, suggesting that the percolation threshold is not inversely proportional to the excluded area, but is apparently linear with its reciprocal, that is,
\begin{equation}
 N_c-N_{c0}\propto A_r^{-1}
 \label{eq:four},
\end{equation}
where $N_{c0}$ is a corrected factor (constant). From linear fitting (adjusted $R^2 = 0.9995$) by Eq. (\ref{eq:four}) to all $N_c$, we obtain $N_{c0} = 0.20(4)$. Listing the corrected product $\left(N_c-N_{c0}\right)A_r$ in Table \ref{tab:table2}, we find it almost retains constant at 3.4 or 3.5 for all $r > 1$. The maximum uncertainty for the prediction of $N_c$ by Eq. (\ref{eq:four}) is \~0.04. This uncertainty is given by the half-width of 99\% confidence interval \cite{Ref29} of the predictions.

\begin{table}
\caption{\label{tab:table3}Coefficients $c_i$ for nonlinear regression of Eq. (\ref{eq:five}) to the simulation data in Table \ref{tab:table2}. The uncertainty is the half-width of the 95\% confidence interval. In this table, all $c_i$ are rounded to 5 decimal places which may be necessary for the prediction of precise $N_c$ by Eq. (\ref{eq:five}), although their uncertainty does not really support such a precision.}
\begin{ruledtabular}
\begin{tabular}{ccc}
\mbox{$i$}& \mbox{$  c_i$} & \mbox{    uncertainty} \\
\hline
1&	5.82930&	0.0243 \\
2&	7.94992&	0.453 \\
3&	-47.48282&	3.44\\
4&	132.30651&	14.0\\
5&	-231.15515&	33.7\\
6&	264.35127&	49.7\\
7&	-190.88090&	44.0\\
8&	78.92223&	21.4\\
9&	-14.20310&	4.43\\
\end{tabular}
\end{ruledtabular}
\end{table}

\begin{table}
\caption{\label{tab:table4}Comparison between percolation thresholds obtained from simulations and calculations through Eq. (\ref{eq:five}) for homogeneous rectangle systems with $r$ other than those in Table \ref{tab:table2}. $\Delta N_c$ is the absolute values of the percolation threshold difference between simulations and calculations.}
\begin{ruledtabular}
\begin{tabular}{cccc}
\mbox{$r$}& \mbox{$N_c$(simulation)} & \mbox{$N_c$(calculation)} & \mbox{$\Delta N_c$} \\
\hline
$1.1$ &	1.078532(21)&	1.078495(58)&	$3.7 \times 10^{-5}$ \\
1.25 &	1.215636(18)&	1.215581(66)&	$5.5 \times 10^{-5}$ \\
2.5 &	2.083711(26)&	2.083720(60)&	$9.0 \times 10^{-6}$ \\
40 &	5.043120(52)&	5.043080(47)&	$4.0 \times 10^{-5}$ \\
200 &	5.504099(69)&	5.504131(56)&	$3.2 \times 10^{-5}$ \\
\end{tabular}
\end{ruledtabular}
\end{table}

In order to gain more precise analysis, we propose a high-order polynomial [Eq. (\ref{eq:five})] to fit the data in Fig. \ref{fig:3}:
\begin{equation}
 N_c=f(s)=\sum\limits_{i=1}^n{c_is^i}=\sum\limits_{i=1}^n{c_i\left(2\pi^{-1}A_r^{-1}\right)^i}
 \label{eq:five},
\end{equation}
where $s=2\pi^{-1}A_r^{-1}$ and $c_i$ are constants. With $n = 9$, a perfect fitting (adjusted $R^2 > 1- 3 \times 10^{-10}$) can be obtained from nonlinear regression \cite{Ref28}. The fitted coefficients $c_i$ are listed in Table \ref{tab:table3}. As an important outcome, Eq. (\ref{eq:five}) can then predict high-precision $N_c$ for any $r$. Here $n = 9$ is the minimum order number required to ensure small uncertainty of $<7\times 10^{-5}$ for all predictions, a comparable level to those of the simulation results in Table \ref{tab:table2}. 

To validate Eq. (\ref{eq:five}), we investigate several more homogeneous rectangle systems with $r$ other than those in Fig. \ref{fig:3} or Table \ref{tab:table2}. The predicted results [by Eq. (\ref{eq:five})] are compared with the simulation results in Table \ref{tab:table4}. Their difference is really less than the predicted uncertainty $7 \times 10^{-5}$. It is concluded that Eq. (\ref{eq:five}) and the maximum uncertainty $7 \times 10^{-5}$ can provide high-precision percolation thresholds for rectangle systems with any aspect ratio $r$. Although Eq. (\ref{eq:five}) is merely based on our simulation results rather than physical deductions, it makes it unnecessary for most future studies to explore the percolation thresholds once more, and should facilitate many studies concerning continuum percolation of rectangles or the layered 2D nanomaterials.

It is worthy of mentioning that Eqs. (\ref{eq:four}) and (\ref{eq:five}) only provide {\it mathematical} fittings to the simulated $N_c$ in Fig. \ref{fig:3}, where the average excluded area $A_r$ is limited within $\left[A_\infty,A_1\right]$ . One may not extrapolate them for $A_r$ outside this region. For example, as $A_r\rightarrow\infty$, it is reasonable to anticipate $N_c\rightarrow0$. However, Eq. (\ref{eq:four}) expects $N_c$ converges to a finite value $N_{c0}$. Note that in heterogeneous systems, the effective average excluded areas may go beyond the region $\left[A_\infty,A_1\right]$, although it is not possible for homogeneous systems. Relevant issues are discussed below in Sec. \ref{sec:IIIC2}.

\begin{figure}
\includegraphics[scale=0.3]{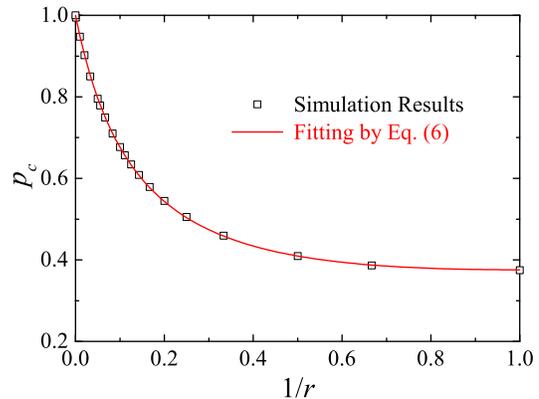}% Here is how to import EPS art
\caption{\label{fig:4}(Color online) The critical remaining area fraction $p_c$ for homogeneous rectangle systems of various aspect ratios (listed in Table \ref{tab:table2}).}
\end{figure}

In addition, Eq. (\ref{eq:five}) should not be the exclusive form to describe the thresholds in rectangle percolation. Other forms may also apply as long as their prediction uncertainty is sufficiently small. Interestingly, Xia {\it et al.} \cite{Ref30} proposed a simple interpolation formula to describe the dependence of critical remaining area fraction $p_c$ on the aspect ratios for random ellipse systems as $p_c=(1+4y)/(19+4y)$, where $y=a/b+b/a$ with $a$ and $b$ being the major and minor semiaxes respectively, and $p_c=\exp(-N_cS)$ with $S$ being the ellipse area. We find the form of this interpolation formula also applies to the rectangle systems in this work despite the different numerical values for the coefficients, that is
\begin{equation}
 p_c=\exp\left(-N_cS\right)=\left(1.28+y\right)/\left(6.73+y\right)
 \label{eq:six},
\end{equation}
where $y = l/w + w/l = r + 1/r$ and for rectangles, $S = lw = l^2/r$. As shown in Fig. \ref{fig:4}, Eq. (\ref{eq:six}) fits the simulation data very well (the maximum prediction uncertainty is about 0.005). However, in this work we mainly discuss the relation between $N_c$ and $A_r$, i.e., Eq. (\ref{eq:five}), which is more important for heterogeneous systems as discussed below. 
\subsection{\label{sec:IIIC}Heterogeneous rectangle systems}	
\begin{figure}
\includegraphics[scale=0.4]{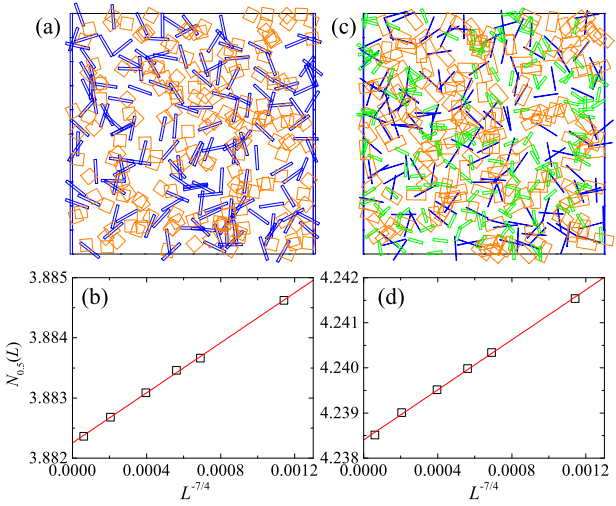}% Here is how to import EPS art
\caption{\label{fig:5}(Color online) Examples of two heterogeneous systems. (a) A percolating realization ($L = 10$) for the binary system of [(1, 0.1), 0.5; (0.5, 0.5), 0.5]. (b) $N_{0.5}(L)$ against $L^{-1-1/v}$ for the binary system in (a). (c) A percolating realization ($L = 10$) for the ternary system of [(1, 0.02), 0.3; (0.8, 0.4), 0.4; (0.6, 0.1), 0.3]. (d) $N_{0.5}(L)$ against $L^{-1-1/v}$ for the ternary system in (c).}
\end{figure}

In this section, we explore the relations between percolation threshold and excluded area for heterogeneous rectangle systems. For simplicity, we denote a rectangle of length $l$ and width $w$ as $(l, w)$. A heterogeneous rectangle system is denoted as $[(l_1, w_1), x_1; (l_2, w_2), x_2; \dots; (l_n, w_n), x_n]$ if it contains $n$ types of rectangles and the number fraction of the $i$th type of rectangles $(l_i, w_i)$ is $x_i$ ($i = 1, 2,\dots, n$ and $\sum\nolimits_{i=1}^n{x_i}=1$). Accordingly, a group of rectangle systems are denoted as ${(l_1, w_1); (l_2, w_2);\dots; (l_n, w_n)}$ which include all systems comprising the $n$ types of rectangles with any combinations of the number fractions.
Some examples for the heterogeneous rectangle systems are shown in Fig. \ref{fig:5}. The Monte Carlo simulation results verifies that Eq.(\ref{eq:three}) still applies to these systems [Figs. \ref{fig:5}(b) and \ref{fig:5}(d)] and high-precision percolation thresholds can also be extracted. In this section, however, we normally use a threshold estimate $N_{0.5}(L)$ (typically $L \geq 64$ and based on $>10^5$ Monte Carlo simulations) as a rough estimation of $N_c$ which, notwithstanding, has already provided sufficient precision (to about three reliable significant figures) for most discussions and enabled us to efficiently establish the percolation threshold-excluded area relation for the heterogeneous systems.

\subsubsection{\label{sec:IIIC1}Generalization of the excluded area theory}
First of all, the average excluded area formula [Eq. (\ref{eq:two})] should be generalized to heterogeneous rectangle systems. Referring to Fig. 2 of Ref. \cite{Ref9}, one can readily find the excluded area between two difference rectangles, $(l_i, w_i)$ and $(l_j, w_j)$, is 
\begin{multline}
%\begin{split}
A_\theta^{ij}\left(l_i,w_i;l_j,w_j\right)=l_iw_i+l_jw_j+\left(l_il_j+w_iw_j\right)\sin\theta \\
 +\left(l_iw_j+w_il_j\right)\cos\theta
%\end{split}
\label{eq:seven},
\end{multline}
where $\theta$ is the angle between the orientations of the two rectangles. If the orientations of the rectangles are uniformly distributed within the region $[-\theta_\mu, \theta_\mu]$, the average excluded area between the different two rectangles is [see Eqs. (6), (7) and (19) in Ref. \cite{Ref9} for more details]
\begin{multline}
%\begin{split}
A^{ij}\left(l_i,w_i;l_j,w_j\right)\\
=l_iw_i+l_jw_j+\left(l_il_j+w_iw_j\right)\left(2\theta_\mu -\sin2\theta_\mu\right)/2\theta_\mu^2\\
 +\left(l_iw_j+w_il_j\right)\left(1-\cos2\theta_\mu\right)/2\theta_\mu^2
%\end{split}
\label{eq:eight}.
\end{multline}
For isotropic ($\theta_\mu = \pi/2$) systems, Eq. (\ref{eq:eight}) becomes
\begin{multline}
%\begin{split}
A^{ij}\left(l_i,w_i;l_j,w_j\right)=l_iw_i+l_jw_j+2\pi^{-1}\left(l_il_j+w_iw_j\right)\\
 +4\pi^{-2}\left(l_iw_j+w_il_j\right)
%\end{split}
\label{eq:nine}.
\end{multline}
Equation (\ref{eq:nine}) suggests $A^{ij}=A^{ji}$ and it degrades to Eq. (\ref{eq:two}) when $i = j$. For a system $[(l_1, w_1), x_1; (l_2, w_2), x_2;\dots;$ $ (l_n, w_n), x_n]$, the effective average exclude area $A_e$ is
\begin{equation}
 A_e=\sum\nolimits_{i=1}^n{\left(x_i\sum\nolimits_{j=1}^n{x_jA^{ij}}\right)}
 \label{eq:ten}.
\end{equation}
For simplicity, in this work, we only explore the $N_c-A_e$ relations for heterogeneous systems with the number of rectangle types $n \le 3$, i.e., the binary ($n = 2$) and ternary ($n = 3$) systems. Nevertheless, the generalization into systems comprising more rectangle types ($n >3$) is anticipated to be straightforward.
\subsubsection{\label{sec:IIIC2}Binary systems}
\begin{figure}
\includegraphics[scale=0.3]{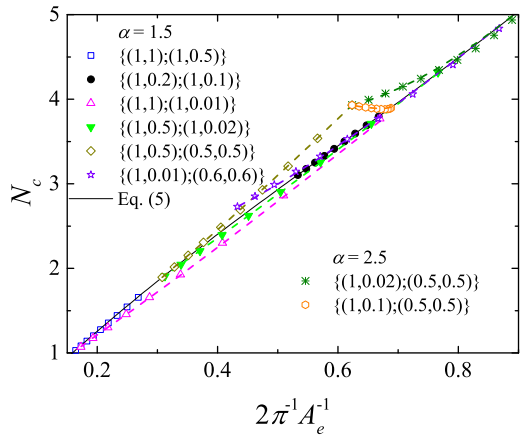}% Here is how to import EPS art
\caption{\label{fig:6}(Color online) Percolation thresholds obtained from simulations (symbols) and calculations (dashed curves) from Eq. (\ref{eq:15}) for weakly and strongly correlated binary systems. For each group of systems, the number fractions $x_i$ varies between 0 and 1 at the step of 0.1. For calculations, only the systems {(1, 0.1); (0.5, 0.5)} and {(1, 0.02); (0.5, 0.5)} are with $\alpha = 2.5$ in Eq. (\ref{eq:15}), and classified as strongly correlated systems. All the others are weakly corrected systems with $\alpha = 1.5$. Note for efficiency, not all $N_c$ in this plot is given by the thermodynamic limit ($L\rightarrow\infty$) values as extrapolated from Eq. (3). For {(1, 1); (1, 0.01)} and {(1, 0.5); (1, 0.02)}, $N_c$ is approximated by $N_{0.5}(L = 64)$, while for {(1, 0.5); (0.5, 0.5)} and {(1, 0.01); (0.6, 0.6)}, $N_c$ is approximated by $N_{0.5}(L = 128)$. Actually, however, with the respect to the discussions of heterogeneous systems in this work, the difference between $N_{0.5}(L \geq 64)$ and the corresponding thermodynamic limit value is negligible [see Figs. \ref{fig:2}, \ref{fig:5}(b) and \ref{fig:5}(d)].}
\end{figure}

We have investigated a series of binary systems, as shown in Fig. \ref{fig:6} where the average excluded areas are calculated from Eq. (\ref{eq:ten}). It is clear in Fig. \ref{fig:6} when the excluded areas are comparable, the systems, no matter whether heterogeneous or homogeneous, have comparable percolation thresholds. It can thereby be confirmed that percolation thresholds are dominated by the average excluded areas. Rigorously, however, slight deviation can be seen between heterogeneous and homogeneous systems. Fortunately, for most binary systems, we have found a general law to describe their precise $N_c-A_e$ relations on the basis of Eq. (\ref{eq:five}).

First, one should note that Eq. (\ref{eq:five}) is correct only when the rectangle systems are measured in units of the rectangle length $l$. Certainly, it is possible that a rectangle system is measured by another unit $l^\prime$ ($l^\prime\not=l$). Under the new measurement, it can be readily established that the values of the percolation threshold and excluded area, denoted as $N_{c,l^\prime}$ and $A_r^\prime$  respectively, relate to those of the old systems as
\begin{equation}
 l^{\prime-2}N_{c,l^\prime}=l^{-2}N_{c,l}=l^{-2}N_c
 \label{eq:eleven},
\end{equation}
and
\begin{equation}
 l^{\prime2}A_r^\prime=l^{2}A_r
 \label{eq:twelve}.
\end{equation}
Then still based on Eq. (\ref{eq:five}), one can predict $N_{c,l^\prime}$ through
\begin{subequations}
\begin{equation}
\begin{split}
N_{c,l^\prime}&=\left(l/l^\prime\right)^{-2}f\left(2\pi^{-1}A_r^{-1}\right)\\
&=\left(l/l^\prime\right)^{-2}f\left[2\pi^{-1}A_r^{\prime-1}\left(l/l^\prime\right)^{2}\right]
 \label{eq:13a},
\end{split}
\end{equation}
In particular, if a homogeneous systems comprising rectangles $(l_i, w_i)$ is measured by  $l^\prime=1$, but $l^\prime\not=l$, Eq. (\ref{eq:13a}) turns to be
\begin{equation}
N_{c,l^\prime}=l_i^{-2}f\left(2\pi^{-1}A_r^{-1}\right)
=l_i^{-2}f\left(2\pi^{-1}A_r^{\prime-1}l_i^{2}\right)
 \label{eq:13b}.
\end{equation}
\end{subequations}
Equation (\ref{eq:13b}) is very useful for our discussion below about heterogeneous systems comprising rectangles of different lengths.

Now we consider a binary system $[(l_1, w_1), x_1;$ $ (l_2, w_2), x_2]$ which is measured by $l^\prime$ (the larger one between $l_1$ and $l_2$). With respect to one type of rectangles $(l_i, w_i)$ ($i$ = 1 or 2), the effective excluded area of the binary system $A_e$ can be regarded as a deviation from that of the homogeneous system $A_{r_i}$ (still measured by $l^\prime$). Note roughly $N_c$ is linear with $1/{A_r}$, as described by Eq. (\ref{eq:four}). Then one can use the linear approximation around $2\pi^{-1}A_{r_i}^{-1}$ as one estimate for the percolation threshold of the binary system, that is
\begin{multline}
N_c^{(i)}\approx l_i^{-2}f\left(2\pi^{-1}A_{r_i}^{-1}l_i^{2}\right) \\
+2\pi^{-1}f^\prime\left(2\pi^{-1}A_{r_i}^{-1}l_i^{2}\right)\left(A_e^{-1}-A_{r_i}^{-1}\right)
 \label{eq:14},
\end{multline}
where $f(s)$ is given by Eq. (\ref{eq:five}) and $f^\prime(s)$ is its first derivative. Note that $N_c^{(1)}$ may differ significantly from $N_c^{(2)}$. Then the actual $N_c$ should result from the correlation between $N_c^{(1)}$ and $N_c^{(2)}$. Obviously, both $N_c^{(1)}$ and $N_c^{(2)}$ should contribute to $N_c$, but their contributions may have unequal weight. It is reasonable to put more weight to the dominant rectangles which in essence contribute more to the system percolation (spanning) when the effect of the fractions ($x_i$) is not considered. In general, the dominant rectangles are those of larger length or of larger width in the case of equal lengths. For convenient discussion, we suppose the dominate rectangles in a binary system are always $(l_1, w_1)$, that is to say, either of the two conditions should be satisfied: (1) $l_1 > l_2$, or (2) $l_1 = l_2$ and $w_1 > w_2$. It is found that $N_c$ of the binary system can be described as
\begin{equation}
N_c=N_c^{(1)}\left(1-x_2^\alpha\right)+N_c^{(2)}x_2^\alpha
 \label{eq:15},
\end{equation}
where $\alpha > 1$ is the correlation exponent. 

\begin{figure}
\includegraphics[scale=0.3]{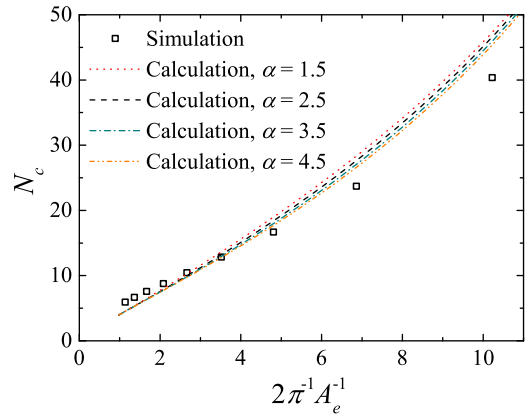}% Here is how to import EPS art
\caption{\label{fig:7}(Color online) Percolation thresholds obtained from simulations (open squares) and calculations (curves) from Eq. (\ref{eq:15}) for a group of ultra-strongly correlated binary systems {(1, 0.01); (0.1, 0.1)}. For the open squares, for left to right, the number fraction $x_1$ increases from 0.1 to 0.9 at a step of 0.1. Note for these systems, $A_e \ll A_\infty$, and the predictions by Eq. (\ref{eq:15}) with all $\alpha$ deviate significantly from the simulations. All $N_c$ is approximated by $N_{0.5}(L = 64)$.}
\end{figure}

In terms of the applicability of Eq. (\ref{eq:15}), we roughly classify the binary systems into three categories: weakly correlated systems (roughly $l_1 < 2l_2$, Fig. \ref{fig:6}), strongly correlated systems ($l_1 \approx 2l_2$ and usually $r_1 \geq 10$, Fig. \ref{fig:6}) and ultra-strongly correlated systems (roughly $l_1 > 2l_2$ and $r_1 \geq 10$, Fig. \ref{fig:7}). As shown in Fig. \ref{fig:7}, $N_c$ of the weakly and strongly correlated systems agrees excellently with Eq. (\ref{eq:15}). The deviation between the simulations and the fittings by Eq. (\ref{eq:five}) is far less than 0.1. More interestingly, all weakly correlated systems share the common correlation exponent $\alpha = 1.5$, while the strongly correlated systems have the common $\alpha = 2.5$. The common exponents $\alpha$ make Eq. (\ref{eq:15}) non-trivial and allow the direct calculation of percolation thresholds for any binary systems in these two categories.

For ultra-strongly correlated systems, however, Eq. (\ref{eq:15}) cannot provide good fitting to the simulations, and evident deviation can be observed (Fig. \ref{fig:7}). One important common feature of weakly and strongly correlated systems is that their $A_e$ (measured by $l^\prime =1$) are still within the region $[A_\infty,A_1]$. However, $A_e$ in the ultra-strongly correlated systems has gone beyond this region. For example, for the system $[(1, 0.01), 0.1; (0.1, 0.1), 0.9]$, $A_e = 0.0623 \ll A_\infty = 0.6366$. Then, the employment of Eq. (\ref{eq:15}) requires the extrapolation of Eq. (\ref{eq:five}) far beyond $[A_\infty,A_1]$. This is a risk as discussed in Sec. \ref{sec:IIIB}. Consequently, Eq. (\ref{eq:15}) may become invalid for ultra-strongly correlated systems, especially when $A_e$ goes far beyond the region.

\begin{table}
\caption{\label{tab:table5}Percolation thresholds $N_c$, effective average exclude area $A_e$, and the critical coverage area $N_cS$ for some ultra-strongly correlated binary systems and homogeneous systems. For ultra-strongly binary systems,  $N_c$ is obtained from Monte Carlo simulations and $S=\sum\nolimits_{i=1}^n{x_il_iw_i}$. For homogeneous systems, $N_c$ and $A_e$ are calculated from Eqs. (\ref{eq:eleven}) and (\ref{eq:twelve}), respectively.}
\begin{ruledtabular}
\begin{tabular}{lccc}
\mbox{System}& \mbox{$N_c$} & \mbox{$A_e$} & \mbox{$N_cS$} \\
\hline
[(1,0.01),0.05;(0.01,0.01),0.95]&	107.5&	0.0040&	0.064\\
Homogeneous (0.05, 0.0167)&	933.4&	0.0041&	0.78 \\
$[(1,0.01),0.1;(0.01,0.01),0.9]$&	53.77&	0.0107&	0.059\\
Homogeneous (0.05, 0.05)&	392.9&	0.0102&	0.98 \\
\end{tabular}
\end{ruledtabular}
\end{table}
These ultra-strongly correlated systems are actually very interesting for both theoretical studies and practical applications. In Table \ref{tab:table5}, we compare some of these systems with the homogeneous systems which have comparable average excluded areas. Surprisingly, these ultra-strongly correlated systems have significantly lower $N_c$, as well as significantly lower critical coverage area $N_cS$, than the homogeneous systems. On one hand, this can be regarded as a breakdown of the excluded area theory which in essence expects comparable excluded areas lead to comparable percolation thresholds. On the other hand, the significantly lower critical coverage area for ultra-strongly correlated systems may explain the experimental observation \cite{Ref16} that the composites of graphene and carbon nanotubes often produce improved performance (over pure graphene films) for transparent conductors which prefer low coverage area (to gain more transmittance) upon percolation. Unfortunately, however, the full understanding of these systems may have to rely on further extensive researches. We may discuss this topic separately in the future.

\subsubsection{\label{sec:IIIC3}Ternary systems}
\begin{figure}
\includegraphics[scale=1]{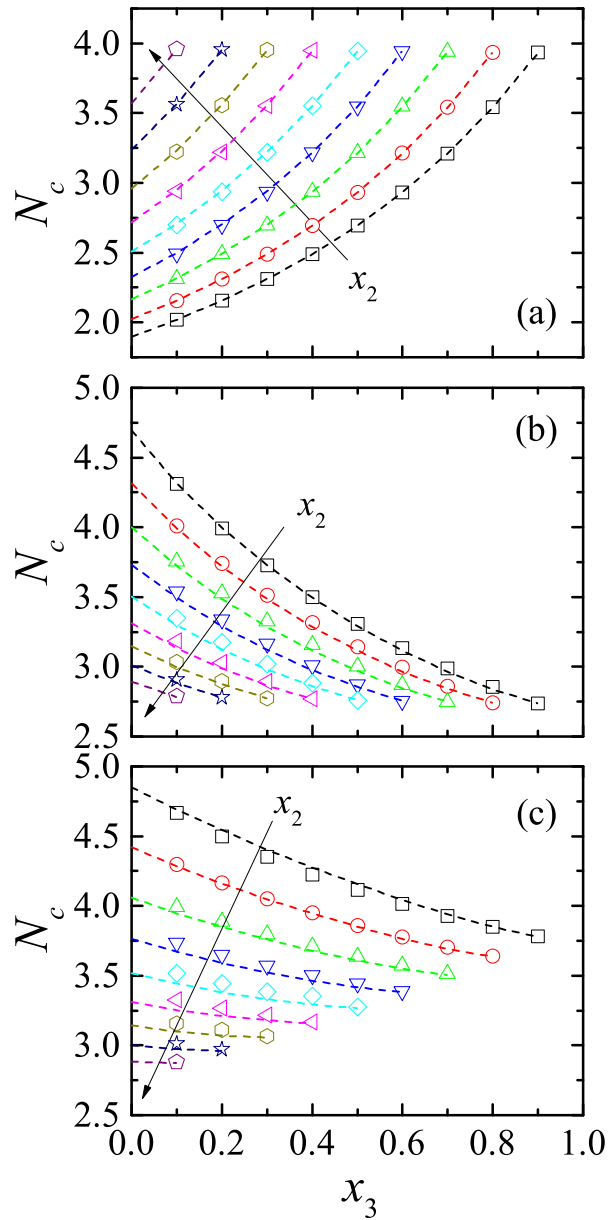}% Here is how to import EPS art
\caption{\label{fig:8}(Color online) Percolation thresholds obtained from simulations (symbols) and calculations (dashed curves) for three groups of ternary systems: (a) {(1, 0.5); (0.6, 0.4); (0.5, 0.5)}, (b) {(1, 0.02); (0.8, 0.4); (0.6, 0.6)} and (c) {(1, 0.01); (0.8, 0.4); (0.5, 0.5)}. All calculations are from Eq. (\ref{eq:16}) with $\alpha=\alpha^\prime=1.5$. In all the plots, along the direction of the arrows, the number fraction $x_2$ increases from 0.1 to 0.9 at a step of 0.1. Note in ternary systems, $x_1 = 1- x_2 - x_3$. All $N_c$ is approximated by $N_{0.5}(L = 64)$.}
\end{figure}

We now extend our discussion to the ternary systems without ultra-strong correlations. It is found that Eq. (\ref{eq:15}) may be generalized to be
\begin{equation}
N_c=N_c^{(3)}x_3^{\alpha^\prime}+\left(1-x_3^{\alpha^\prime}\right)\left[N_c^{(2)}\hat{x}_2^\alpha+\left(1-\hat{x}_2^\alpha\right)N_c^{(1)}\right]
 \label{eq:16},
\end{equation}
where $\hat{x}_2=x_2/(x_1+x_2)$, $\alpha$ and $\alpha^\prime$ are appropriate correlation exponents, and $N_c^{(i)}$ are still defined by Eq. (\ref{eq:14}). Note in Eq. (\ref{eq:16}), we suppose the rectangles have already been sorted so that if $i < j$, $l_i > l_j$ or $w_i > w_j$ when $l_i = l_j$. It might be difficult to determine the exact values for $\alpha$ and $\alpha^\prime$ in ternary systems because the correlations are among three types of rectangles. In most case, however, one may simply take $\alpha=\alpha^\prime=1.5$ at the expense of possibly a little higher deviation. Figure \ref{fig:8} compares the simulation and calculation results for three groups of ternary rectangle systems, {(1, 0.5); (0.6, 0.4); (0.5, 0.5)}, {(1, 0.02); (0.8, 0.4); (0.6, 0.6)} and {(1, 0.01); (0.8, 0.4); (0.5, 0.5)}. Again, calculations are in excellent agreements with simulations. The systems {(1, 0.5); (0.6, 0.4); (0.5, 0.5)} and {(1, 0.02); (0.8, 0.4); (0.6, 0.6)} only involve weak correlations and ideal predictions (maximum deviation < 0.05) are achieved by Eq. (\ref{eq:16}) [Figs. \ref{fig:8}(a) and \ref{fig:8}(b)], while the systems {(1, 0.01); (0.8, 0.4); (0.5, 0.5)} involve strong correlations [possibly between the rectangles (1, 0.01) and (0.5, 0.5)] and the predictions have a little higher deviation (maximum deviation $\sim 0.07$) from the simulations [Fig. \ref{fig:8}(c)].

\section{\label{sec:IV}CONCLUSIONS}
In summary, integrating the Cohen-Sutherland algorithm into our previously-introduced high-efficiency algorithm for continuum percolation, we have developed an efficient Monte Caro algorithm for both homogeneous and heterogeneous rectangle systems. This algorithm enables us to conduct extensive simulations and report high-precision percolation thresholds for a variety of rectangle systems. For homogeneous systems, in terms of the excluded area reciprocals, this work produces a high-order polynomial [Eq. (\ref{eq:five})] which can predict high-precision percolation thresholds (uncertainty $< 7 \times 10^{-5}$) for any rectangle aspect ratios. On the basis of this polynomial, a general and simple relation [Eqs. (\ref{eq:15}) and (\ref{eq:16})] is discovered which can thereby be used to predict precise percolation thresholds for heterogeneous rectangle systems without ultra-strong correlations. Therefore, this work verifies the traditional excluded area theory and extends it into heterogeneous systems. We expect our algorithm and the attained results are useful for many practical applications and theoretical studies concerning 2D layered nanomaterials and various rectangle continuum percolations.

\section*{ACKNOWLEDGMENTS}
Most simulations were performed on the computer cluster “Ferlin” at PDC Center in KTH, and some were on resources provided by the Swedish National Infrastructure for Computing (SNIC) at C3SE. We acknowledge the financial support by the Advanced Investigator Grant (OSIRIS, Grant No. 228229) from the European Research Council and the 7th Framework Program (NANOFUNCTION, NoE 257375) from the European Commission.


\begin{thebibliography}{99}

 \bibitem{Ref1}
S. D. Sarma, S. Adam, E. H. Hwang and E. Rossi, Rev. Mod. Phys. {\bf 83}, 407 (2011).
\bibitem{Ref2}
K.-G. Zhou, N.-N. Mao, H.-X. Wang, Y. Peng, and H.-L. Zhang, Angew. Chem. Int. Ed. {\bf 50}, 10839 (2011). 
\bibitem{Ref3}
J. N. Coleman, et al., Science {\bf 331}, 568 (2011).
\bibitem{Ref4}
D. Golberg, Y. Bando , Y. Huang , T. Terao , M. Mitome , C. Tang and C. Zhi, ACS Nano {\bf 4}, 2979 (2010).
\bibitem{Ref5}
G. Eda and M. Chhowalla, Nano Lett. {\bf 9}, 814 (2009).
\bibitem{Ref6}
M. Hempel, D. Nezich, J. Kong, and M. Hofmann, Nano Lett. {\bf 12}, 5714 (2012).
\bibitem{Ref7}
J. Li, F. Ye, S. Vaziri, M. Muhammed, M. C. Lemme and M. \"Ostling, Adv. Mater. DOI: 10.1002/adma.201300361 (2013).
\bibitem{Ref8}
J. Hicks, A. Behnam, and A. Ural, Appl. Phys. Lett. {\bf 95}, 213103 (2009).
\bibitem{Ref9}
I. Balberg, C. H. Anderson, S. Alexander and N. Wagner, Phys. Rev. B {\bf 30}, 3933 (1984).
\bibitem{Ref10}
I. Balberg, Phys. Rev. B {\bf 31}, 4053(R) (1985).
\bibitem{Ref11}
J. A. Quintanilla and R. M. Ziff, Phys. Rev. E {\bf 76}, 051115 (2007).
\bibitem{Ref12}
S. Mertens and C. Moore, Phys. Rev. E {\bf 86}, 061109 (2012).
\bibitem{Ref13}
D. R. Baker, G. Paul, S. Sreenivasan, and H. E. Stanley, Phys. Rev. E {\bf 66}, 046136 (2002).
\bibitem{Ref14}
J. Li and S.-L. Zhang, Phys. Rev. E {\bf 80}, 040104(R) (2009).
\bibitem{Ref15}
Z. Néda, R. Florian, and Y. Brechet, Phys. Rev. E {\bf 59}, 3717 (1999).
\bibitem{Ref16}
V. C. Tung, L.-M. Chen, M. J. Allen, J. K. Wassei, K. Nelson, R. B. Kaner, and Y. Yang, Nano Lett. {\bf 9}, 1949 (2009).
\bibitem{Ref17}
G. Cunningham, M. Lotya, N. McEvoy, G. S. Duesberg, P. van der Schoot, and J. N. Coleman, Nanoscale {\bf 4}, 6260 (2012).
\bibitem{Ref18}
M. E. J. Newman and R. M. Ziff, Phys. Rev. Lett. {\bf 85}, 4104 (2000); Phys. Rev. E 64, 016706 (2001).
\bibitem{Ref19}
T. Vicsek and J. Kertész, J. Phys. A {\bf 14}, L31 (1981).
\bibitem{Ref20}
J. D. Foley, {\it Computer graphics: principles and practice}, 2nd ed., p. 113 (Addison-Wesley, 1996); http://en.wikipedia.org/wiki/Line\_clipping
\bibitem{Ref21}
R. M. Ziff and M. E. J. Newman, Phys. Rev. E {\bf 66}, 016129 (2002).
\bibitem{Ref22}
J. Li and M. \"Ostling, Phys. Rev. E {\bf 86}, 040105(R) (2012).
\bibitem{Ref23}
When preparing this manuscript, we noticed a preprint \cite{Ref12} employed almost the same algorithm (combination of Newman-Ziff and subcell) but applied to systems with periodic boundary conditions in both horizontal and vertical directions, and used accordingly threshold estimates based on wrapping probability \cite{Ref18}, instead of spanning probability for FBCs in this work, to extract more precise values for percolation thresholds in some other continuum systems. The technique may also apply to rectangle systems in this work. Nevertheless, our attained percolation thresholds should already be sufficiently precise for most practical and theoretical applications.
\bibitem{Ref24}
J. Li, B. Ray, M. A. Alam, and M. \"Ostling, Phys. Rev. E {\bf 85}, 021109 (2012).
\bibitem{Ref25}
G. E. Pike and C. H. Seager, Phys. Rev. B {\bf 10}, 1421 (1974).
\bibitem{Ref26}
Y.-D. Liang and B. A. Barsky, ACM Trans. Graph. {\bf 3}, 1 (1984).
\bibitem{Ref27}
http://en.wikipedia.org/wiki/Coefficient\_of\_determination
\bibitem{Ref28}
http://www.mathworks.se/help/stats/nlinfit.html 
\bibitem{Ref29}
http://www.mathworks.se/help/stats/nlpredci.html 
\bibitem{Ref30}
W. Xia and M. F. Thorpe, Phys. Rev. A {\bf 38}, 2650 (1988).

\end{thebibliography}
\end{document}

%\preprint{APS/123-QED}

\title{Erratum: Percolation thresholds of two-dimensional continuum systems of rectangles [Phys. Rev. E 88, 012101 (2013)]}% Force line breaks with \\
%%\thanks{A footnote to the article title}%

\author{Jiantong Li}
%% \altaffiliation[Also at ]{Physics Department, XYZ University.}%Lines break automatically or can be forced with \\
 %\email{jiantong@kth.se}
\author{Mikael \"{O}stling}%
 %%\email{Second.Author@institution.edu}
\affiliation{
KTH Royal Institute of Technology, School of Information and Communication Technology, Electrum 229, SE-164 40 Kista, Sweden
 %%This line break forced with \textbackslash\textbackslash
}

%%\collaboration{MUSO Collaboration}%\noaffiliation

\date{\today}% It is always \today, today,

\pacs{64.60.ah}% PACS, the Physics and Astronomy
\maketitle

%\tableofcontents
With the recent publication by Chatterjee \cite{Chatterjee}, we notice that the original formula, given as Eq. (18) in Ref. \cite{Balberg}, for the excluded area for two identical rectangles is incorrect. The correct form is given as Eq. (1) in Ref. \cite{Chatterjee}, as
\begin{equation}
 A_\theta=(l^2+w^2)|\sin\theta|+2lw(1+|\cos\theta|)
 \label{eq:one},
\end{equation}
where {\it l} and  {\it w} are the length and width of the rectangles, and $\theta$ is the angle between their longitudinal axes. Note that the absolute value signs are missing for $\sin \theta$  and $\cos \theta$  in Eq. (18) of Ref. \cite{Balberg}, and Eq. [7] of our paper \cite{Li1}. For isotropic rectangle systems, the average excluded area over uniformly-distributed $\theta$ is given by Eq. (9) in Ref. \cite{Chatterjee}, as
\begin{equation}
 A_r=2\pi^{-1} (l^2+w^2)+2(1+2\pi^{-1})lw
 \label{eq:two}.
\end{equation}
In general, the average excluded area for two randomly-oriented unequal rectangles {\it i} and {\it j} is then given as
\begin{equation}
 A_{ij}=(l_iw_i+l_jw_j)+2\pi^{-1}(l_il_j+w_iw_j+l_iw_j+l_jw_i)
 \label{eq:three},
\end{equation}
where $l_i$ and $w_i$ are the length and width of the rectangle {\it i}, respectively.

Equations (\ref{eq:two}) and (\ref{eq:three}) are consistent with the general form \cite{Mark} for the average excluded area of any two randomly-oriented convex shapes {\it i} and {\it j}, as
\begin{equation}
 A_{ij}=S_i+S_j+C_iC_j/2\pi
 \label{eq:four},
\end{equation}
where $S_i$ and $C_i$ are the area and perimeter of the convex shape {\it i}, respectively. For rectangles, $S_i=l_iw_i$ and $C_i=2(l_i+w_i)$, and Eq. (\ref{eq:four}) turns to be Eq. (\ref{eq:three}) or Eq. (\ref{eq:two}) when $i = j$. 

\begin{figure}
\includegraphics[scale=0.25]{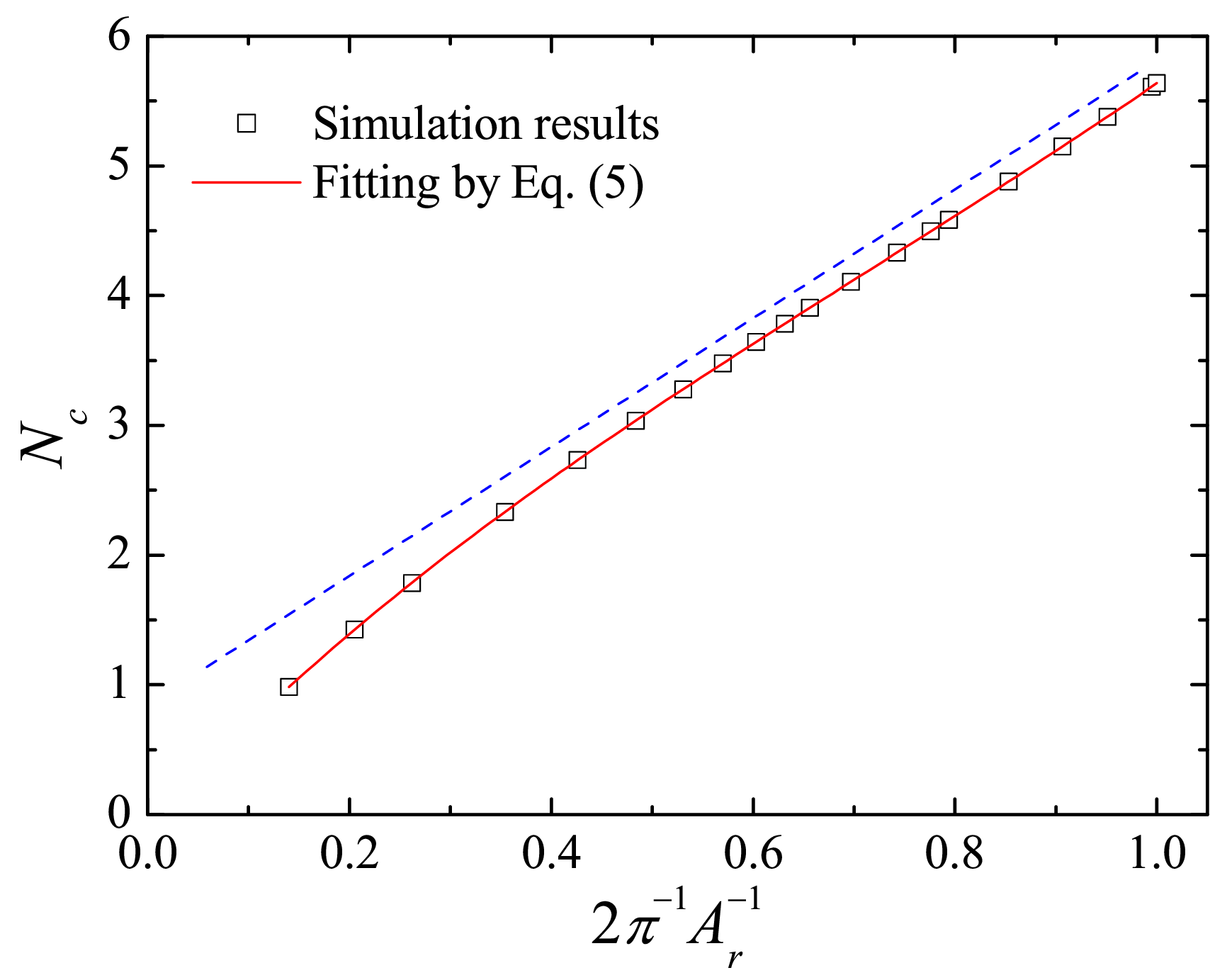}% Here is how to import EPS art
\caption{\label{fig:E1}(Color online) Plots of $N_c$ against {\it s} for homogeneous rectangle systems with aspect ratio $r = l/w$ ranging from 1 to $\infty$. Note $A_r$ is given by Eq. \ref{eq:two} with $l = 1$. The (red) solid curve is the fitting by Eq. (5) in \cite{Li1} with coefficients $c_i$ listed in Table \ref{tab:table1}. The (blue) dashed line is guide to eyes to show the nonlinearity of the $N_c$ - {\it s} relation.}
\end{figure}

In the paper \cite{Li1}, we derived incorrect formulas for the average excluded area of overlapping rectangles, including Eqs. (2) and (9) in \cite{Li1}. They should influence most discussion in \cite{Li1}. Here we plot in Fig. \ref{fig:E1} the correct $N_c$ - {\it s} $(s=2\pi^{-1}A_r^{-1})$ relation for homogeneous rectangle systems, which was originally plotted in Fig. 3 of Ref. \cite{Li1} with the incorrect $A_r$ given by Eq. (2) in \cite{Li1}. Still, with {\it n} = 10, Eq. (5) in \cite{Li1} gives perfect fitting (adjusted $R^2 \approx 1 - 2\times10^{-10}$) to the correct $N_c$ - {\it s} relation of rectangle systems. With the coefficients $c_i$ listed in Table \ref{tab:table1}, Eq. (5) in \cite{Li1} can be used to predict the percolation threshold of any homogeneous systems of rectangle aspect ratio ranging from 1 to $\infty$, with the maximum uncertainty $<6\times 10^5$.

\begin{table}
\caption{\label{tab:table1}Coefficients $c_i$ for nonlinear regression of Eq. (5) in \cite{Li1} to the simulation data plotted in Fig. \ref{fig:E1}. The decimal places and uncertainty of $c_i$ have the same meaning as in Table III in \cite{Li1}. }
\begin{ruledtabular}
\begin{tabular}{ccc}
{\it i}&\mbox{$c_i$}&\mbox{Uncertainty}\\
\hline
1&6.24650&0.036\\
2&14.81660&0.80\\
3&-104.25949&7.28\\
4&	362.02857&	36.4\\
5&	-818.38082&	110.8\\
6&	1256.78569&	214.5\\
7&	-1296.23424&	265.4\\
8&	858.32414&	203.4\\
9&	-329.26745&	88.0\\
10&	55.57775&	16.4\\
\end{tabular}
\end{ruledtabular}
\end{table}

However, as shown in Fig. \ref{fig:E1}, $N_c$ - {\it s} relation does not exhibit good linearity. This prevents Eq. (15) in \cite{Li1} from precisely predicting percolation threshold for most heterogeneous rectangle systems. To attain linear $N_c$ - {\it s} relation and restore the predictability of Eq. (15) in \cite{Li1}, we propose a general modified form for excluded area as
\begin{equation}
 \tilde{A}_{ij}=A_{ij}+\delta_ir_i^{-1}+\delta_jr_j^{-1}
 \label{eq:five},
\end{equation}
where $r_i=l_i/w_i$ and the modification factors $\delta_i$ are constants (independent of $r_i$). Note that the modification aims to obtain linear $N_c$ - {\it s} relation for homogeneous rectangle systems ($r_i=r_j$). In this sense, the incorrect formulas for exclude area of rectangles in \cite{Li1} happen to be acceptable forms (as shown in Fig. 3 in \cite{Li1}) for the modified excluded area   $\tilde{A}_{ij}$ with $\delta_i=\delta_j=l_il_j(4\pi^{-2}-2\pi^{-1})$. Actually, in our recent work \cite{Li2}, similar modification of excluded area is employed in order to precisely predict percolation threshold for heterogeneous systems comprising unequal ellipses.

Consequently, all the excluded area in \cite{Li1} should read as the modified excluded area as defined by Eq. \ref{eq:five} in this Erratum, except for Eqs. (7) and (8) in \cite{Li1} which discuss about the true excluded areas but should be replaced with Eqs. (18) and (26) in Ref. \cite{Chatterjee}, respectively. With this correction, all the results and conclusions in \cite{Li1} remain unaltered. However, the readers interested in the dependence of percolation threshold on the true excluded area in rectangle systems may refer to this Erratum and Ref. \cite{Chatterjee}.

We are grateful to Prof. Avik P. Chatterjee for kindly pointing out the errors in \cite{Li1}.